# GRID INFORMATION SECURITY FUNCTIONAL REQUIREMENT

## FULFILLING INFORMATION SECURITY OF A SMART GRID SYSTEM

Amy Poh Ai Ling[1] and Mukaidono Masao[2]

[1] Department of Mathematical Modeling, Analysis & Simulation, Graduate School of Advanced Mathematical Sciences, Meiji University, Kawasaki, Japan
amypoh@meiji.ac.jp
[2] Department of Computer Science, School of Science and Technology, Meiji University, Kawasaki, Japan
masao@cs.meiji.ac.jp

### *ABSTRACT*

*This paper describes the background of smart information infrastructure and the needs for smart grid information security. It introduces the conceptual analysis to the methodology with the application of hermeneutic circle and information security functional requirement identification. Information security for the grid market cover matters includes automation and communications industry that affects the operation of electric power systems and the functioning of the utilities that manage them and its awareness of this information infrastructure has become critical to the reliability of the power system. Community benefits from of cost savings, flexibility and deployment along with the establishment of wireless communications. However, concern revolves around the security protections for easily accessible devices such as the smart meter and the related communications hardware. On the other hand, the changing points between traditional versus smart grid networking trend and the information security importance on the communication field reflects the criticality of grid information security functional requirement identification. The goal of this paper is to identify the functional requirement and relate its significance addresses to the consumer requirement of an information security of a smart grid. Vulnerabilities may bring forth possibility for an attacker to penetrate a network, make headway admission to control software, alter it to load conditions that destabilize the grid in unpredictable ways. Focusing on the grid information security functional requirement is stepping ahead in developing consumer trust and satisfaction toward smart grid completeness.*

### *KEYWORDS*

*Functional Requirement, Smart Grid, Information Security, Networking Trend, Consumer Requirement*

# 1. INTRODUCTION

Information technology advancement paves way to daily conveniences. Today, we simply can reach someone we know just with a touch on the mobile phone, makes the world become thence small. Distance is no longer a barrier for communication. Likewise, in the nearest future all electronic gadgets could be controlled via programmed interface software. The consumption of electricity at home can be monitored with an access password provided. A culture encroachment happens through the interplay of technology and everyday life. Traditional distribution network management of power supply and loads has been conceived as an independent process. However,this traditional approach is changing bit by bit by an increasing number of distributed channels. The distribution channel covers the means of renewable energy resources. This will heighten active energy resources like loads, storages and plug-in hybrid vehicles. The emerging of smart grid pushes the market towards a drastically increase in the demand of smart supply of energy flow. Grid optimization goes on for system reliability, operational efficiency and asset utilization and protection. Smart grid uses two-way communication systems for better monitoring towards lower energy consumption. In short, we must ensure that the integrators consider on security when combining devices system-wide. The satisfaction of a consumer depends on how strong the information security on the grid can be secured.

# 2. REVIEW OF LITERATURE

The rising of real incomes and the increasing in abundant of consumer goods have pricked demand for greater consumption on electronic gadgets in most modernized countries. Modeling lifestyle effects on energy demand [1] study that the increase of societal energy consumption influenced by three main items: technical efficiency, lifestyles and socio-cultural factors. Technological transfer phenomenon is often seen as a crucial part contributes to the solutions of environmental highlights. However, when technological change is seen from the perspective of everyday life, this image becomes more complex [2].

## 2.1. Grid Information Technology

Electricity distribution networks create a market place for small-scale power producers and for customers - users of electricity. At the core of all smart grid definition is advanced metering infrastructure (AMI). This refers to smart meters and an accompanying communications network that allows two-way communication between the provider of electricity and the meter. With this capability, providers have access to real-time information on the electricity consumption of each customer. Smart metering technology will be the foundation of any Smart Grid strategy, as it provides the platform for offering digital-era services to consumers [3]. To incorporate intermittent energy resources, a category which renewable energy falls into, electricity networks will have to become „smarter grids‟ with integrated communication systems and real time balancing between supply, demand, and storage [4]. One of the greatest challenges for future electricity grids lay on the demand side response and creating a system that can shift peak demand, at the same time as being socially acceptable. This is a key issue as major behavioural changes are necessary to modify energy usage patterns and the current demand curve, likely to be facilitated by suitable user-friendly technology platforms.

AMI technologies play the role of increasing the reliability of the electric power delivery system and bettering customer service [5]. These technologies rely on two-way communications systems either lower capacity or speed narrowband options such as fixed wireless radio frequency (RF) or power line carrier (PLC) or broadband options such as broadband over power lines (BPL) or wireless networks.

For network operation purposes, more accurate real-time state estimation of the whole network gives information on voltages, loads, losses and stressing of components, and also makes it possible to optimize, e.g. network topology, voltage control, and load control actions [6]. The grid itself may be brought off more expeditiously with the power to control loads [7] and the typically higher information quality, followed by the consequences of increased data accumulation throughout the entire system. Grid technology provides the chance of a simple and transparent access to different information sources. A data grid can be interpreted as the consolidation of different data managing systems furnishing the user with data, information and knowledge [8].

### 2.2. Needs for Smart Grid Information Security

A number of smart grid information security requirements and regulations are available online [9] [10] [11] [12], although those guidelines are a significant step in securing the smart grid, but they do not fully address potential vulnerabilities that can emerge. One of the strongest arguments made for securing smart meters is that consumers will have physical [13] and potentially logical access to the smart meters. Security is generally described in terms of availability, integrity, and confidentiality. Cyber systems are vulnerable to worms, viruses, denial-of-service attacks, malware, phishing, and user errors that compromise integrity and availability [14].

It is important to avoid baring any potential security risk. As a matter of fact, grid security technologies have been so far designed and deployed as a middleware layer add-on [15], independently at each tier. The need to protect privacy and security of priceless data over the grid is fueling even more need for common security evaluation criteria. In brief, [16] information security professionals need to be aware that the workings of the most basic IT resource of electricity supply is changing in a manner that introduces a far larger and remotely addressable attack surface combined with the tempting opportunity for mischief and monetary gain.

## 3. RESEARCH METHODOLOGY

Conceptual analysis was adapted as methodology with the application of hermeneutic circle. Reason being hermeneutic circle is an interpretive [17] and a conceptual-analytical research method [18]. This approach is commonly applied by philosophers and theologians in reviewing something that is not explicitly present in it [19]. This study discovers the assumptions of different information security efforts where hermeneutic circle become a natural research methodological choice.

The concept of this paper's methodology is exhibited in Figure 1. It refers to the idea that one understands of the text as a whole is established by reference to the individual parts and one understands of each individual part by reference to the whole [20].

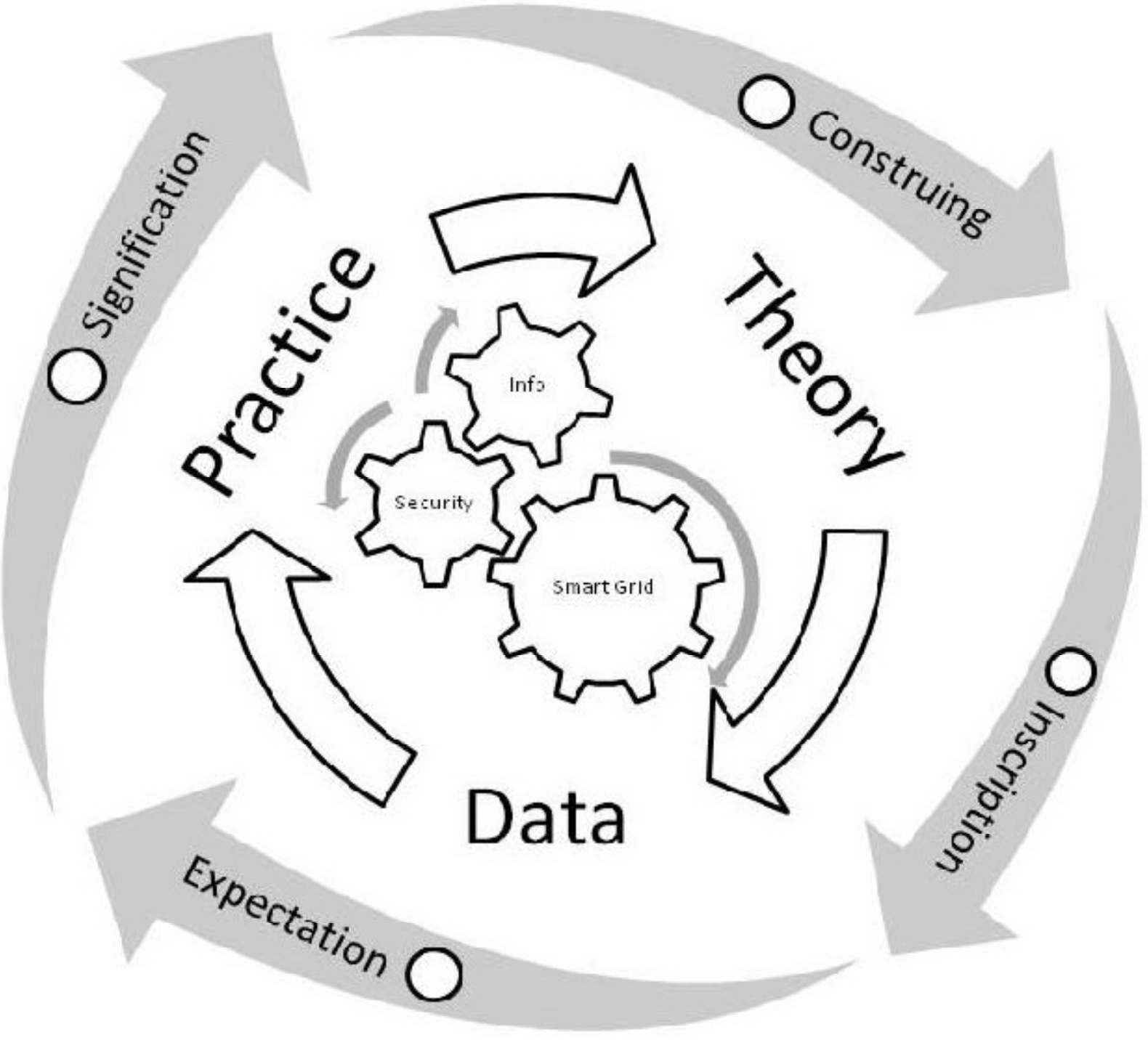

Figure 1. Methodology - Hermeneutic Circle

The approach started with the construing of adducing and explaining the meaning of Smart Grid and Grid Information Technology. This is followed by the inscription of the need for Information Security of a Smart Grid System via literature review. Then the functional requirement serves to achieve consumer requirement expectation assumption are developed. The efforts are then lifted with the significance study carried out. Note that the four main stages for this study were supported by three main elements: theory, data and practice that serve as strong reference for the sources obtained. Information, security and smart grid are the main keywords contributed to the process of literature review to make sure the sourcing does not bump-out from the topic set.

Meanwhile research is a process of collecting; analyzing and interpreting information to answer questions, the research process is similar to undertaking a journey. A knowledge base of research methodology plays a crucial role to kick start a smooth study. Figure 2 displays the practical steps undertaken to complete this paper.

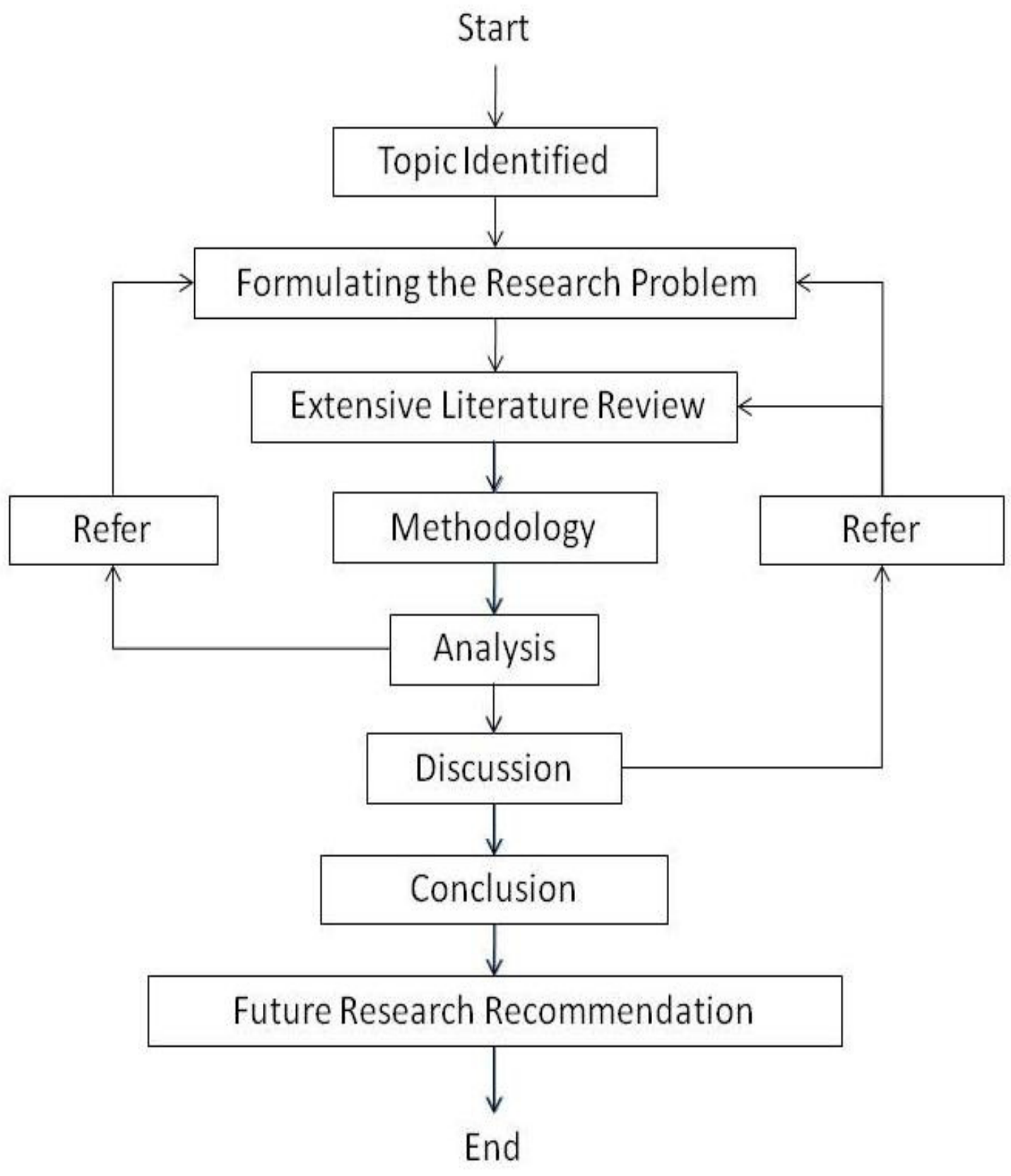

Figure 2. Research Process

It started with the identification of the topic followed by formulating research problem and extensive literature review carefully laid out. Methodology was then developed to direct the path of research process, where analysis and discussion took place. A solid conclusion [21] and future research recommendation was written to complete this paper.

# 4. NETWORKING TREND RELATED TO INFORMATION SECURITY

Smart grid is already leveraging benefits of networking. Its technologies may even integrate with social networking sites [22], as well as associated security risks. This section examines network security, trust, and privacy concerns with regard to networking sites among consumers. Personal data from profiles may also leak outside the network when search engines index them [23]. With the changes of smart grid and other in the electricity utility industries, new demands on the telecom networks were generated [24]. Smart meters as electricity transcription communicate with residential consumer used at least hourly. It transmits at least daily in order to inform more sophisticated time-of-use, real time or other billing structure. Dominant current utility architecture involves one-way flow of electricity from a small number of generation facilities to a large number of consumers, data networks formerly served a predominantly one way flow of information from a small number of mainframe computers to a large number of dumb terminals.

The networking trends are transforming from its traditional flow to an advanced smart grid style of flow, spurring a positive change of networking trends towards a more distributed, digitized and multiple source of supply. Smart Grid Strategy Smart Grid is the nature and compatible use of alternative energy, it needs to create an open computing system and the establishment of a shared information model, based on the data integration system, optimizes network operation and management [25].

Table 1. Networking Trend

positive transformation

| Networking Trends | Traditional | Smart Grid | Communication | Information Security |
|---|---|---|---|---|
| Info Source | Centralized | Distributed | On the grid (Opened computing) | Important |
| Traffic Flow | One-Way | Omni Directional | On the grid (Opened computing) | Important |
| Standards | Vendor-proprietary | Open | On the grid (Opened computing) | Important |
| Coding | Analog | Digital | On the grid (Coding) | Important |
| Control | Private | Public | On the grid (Opened computing) | Important |
| Authority | Single-party | Multi-party | On the grid (Mass computing) | Important |
| Energy Service provider | Single | Multiple | On the grid (Opened computing) | Important |

Resources can be identified and employed in the most effective manner when the environment is fully connected with suitable mechanisms for sharing information, ensuing great positive transformation of electricity service quality and value for both utility provider and end user. As the enabler for the grids, the communications network moulds the base for smart grid towards a qualitative change and a revitalized power utility. The transformation of network flows was portraying in Table 1.

Advanced technologies in physical security and storage capability in a network will produce positive smart grid transformation from its traditional practice. This positive transformation happened on networking trends, such as info source, traffic flow, standards, coding, control, authority and energy service provider. It aspires to assure that households have more proficient and effective control over their energy consumption besides proactively making a reduction on their energy usage.

Constructing smart grids involves positive transformation from traditional electricity networks to new technology system by adding smart sensors, back-end IT systems, smart meters and communications networks. All communication will be on the grid supported by coding system, open computing and mass computing system. The point is, when communication is exposed to open and mass computing, the information security account is considered being enormously important, this leads to the thought on the crucial of information security functional requirement.

# 5. AWARENESS OF GRID INFORMATION SECURITY FUNCTIONAL REQUIREMENT

Smart grid is a "system of systems" [26] where it necessitated the evolution of technology uniformly across the utilities and simpatico with multitudinous of devices. AMI systems capture and process data at the meter [27], furnish information to utilities and consumers. It is usually the case that AMI systems pick-a-back on a variety of wireless systems. While there are other critical questions regarding the role that incumbent"s play in deploying technology that will enable more efficient energy markets [28], smart grid transforms the electricity network into the information age with its digital technology [29].

The underpinning of situational awareness on grid information security is to identify adversaries, estimate impact of attacks, evaluate risks, understand situations and make accurate decisions on how to protect the grid promptly and efficiently. In this paper, situational awareness in grid information security functional requirement is investigated. A good security system for smart grids requires it to be thorough.

It brings the fact that security capacities must be superimposed such that defense mechanisms have sufficient points to detect and mitigate breaches. These capacities should be integral to all segments of the smart grid system features and comply with the full set of grid information security functional requirements.

# 6. INFORMATION SECURITY FUNCTIONAL REQUIREMENT IDENTIFICATION

## 6.1. Identification Process

In the process of identifying information security functional requirement, the significant relationship to consumer requirement is taken into consideration as how it would impact consumer trust and satisfaction. It refers to the consumer requirement developed [30].

Sixteen functional requirements were discussed with the reference to the sixteen consumer requirements identified: info access limitation, data authenticity, data and backup recovery, device and system configuration information, personal key exchange, trusted network, interoperability and security, gap analysis, reliable data storage system, cyber security guidelines, law enforcement, improved wireless technology, controlled power consumption, protect secret, cryptographic protocols and encryption policies, as in Table 2.

Table 2. Information Security Functional Requirement Identification

| Consumer Requirements | Functional Requirement | Discussion |
|---|---|---|
| Confidentiality | Info access limitation | The purpose for which business information is required should be identified before the information is collected and the information collected should be limited to that necessary for the identified purpose. |
| Integrity | Data authenticity | A loss of integrity is the unauthorized modification or destruction of information. |
| Availability | Data and backup recovery | The disaster recovery plan defines the roles and responsibilities and identifies the critical information technology application programs, operating systems, personnel, data files, and time frames needed to ensure high availability and system reliability based on the business impact analysis. |
| Reliable device level | Device and system configuration information | Based on data sensitivity requirements for the information system. |
| Cryptography and key management | Personal key exchange | LRA (Local Registration Authority) identifies and registers end-users, service providers and other independent TTPs securely connected to a CA (Certification Authority). |
| Reliable systems level | Trusted network | The common and unique technical requirements should be allocated to each smart grid system and not necessarily to every component within a system, as the focus is on security at the system level. |
| Networking issues | Interoperability and security | Strong logical separation of network traffic must be achieved using appropriate networking protocols, security tools, and defense-in-depth architecture. |
| Strategic support | Gap analysis | Gap analysis assessment can leverage existing data management and information systems security efforts. |
| Quality assurance | Reliable data storage system | End-user is interested for grid services that are hosted on "high-end" resources including expensive equipment and data storage systems, and which are connected via reliable and high-speed networks. |
| Tactical oversight monitoring system | Cybersecurity guidelines | Need maintenance of a robust and reliable system of oversight, particularly of information systems that support intelligence. |
| Privacy concern | Law enforcement | The concern exists that the prevalence of granular energy data could lead to actions on the part of law enforcement. |
| High bandwidth of communications channels | Improved wireless technology | Technical expertise in both metering and low bandwidth networks are needed to successfully implement enhanced security within the resource. |
| Microprocessor performs memory and compute capabilities | Controlled power consumption | Power consumption has a relation with the program running on a microprocessor. |
| Security in wireless media | Protect secret | Certain data and information are classified secret and need protection. |
| Mature or proprietary protocols | Cryptographic protocols | Cryptographic protocols are used to implement security services. |
| Facilities misuse prevention | Encryption policies | Encryption policies like encrypting sensitive data that is either at rest (databases) on in motion (emails, instant messages, and portable devices). |

1. *Info access limitation*: Receptiveness is necessitated to alleviate public involvement in assessing justifications for technologies, systems or services, identifying purposes of collection, facilitating access and correction by the concerned party, and serve in ascertaining the principles observed [31]. In an open system, all form of information should be protected by appropriate controls against unauthorized access, or alteration, disclosure or destruction and against accidental loss or destruction, and eliminates the need to access customer property.

2. *Data authenticity*: Security is a consolidative concept. It covers availability, authenticity, confidentiality and integrity [32]. Security guards against unconventional information modification or destruction [33], involves ensuring information non-repudiation and authenticity is mandatory [34]. Whenever an integrity or authenticity problem is observed, the system must ensure that data is not exploited.

3. *Data and backup recovery*: The system must always managing, deploying and furbishing up a technology or solution target to maximize the benefits of systems and technology which facilitate to control IT risk [35]. Procedures were developed aiming to render restoration [36], backup, offsite storage and disaster recovery consistent with the entity‟s defined system availability and associated security policies.

4. *Device and system configuration information*: Information relating to the internal functioning of a computer system or network [37], including but not bounded to network and device addresses, device and system configuration information need an average limits of security protect. Taking into account today‟s menace environment, blended with the heightening interoperability and openness, a solid system requires incorporation of assorted security measures. Important security items would be access control, device and application authentication, layoff and fail over for extended operation, encryption for secrecy and escape of sensitive data and information [38]. In a configuration system, it is important to ascertain that the devices are firm and are ready to defend themselves from attacked by firmware updates, not easily swapped out by a rogue one or highjacked by a spoofed remote device.

5. *Personal key exchange*: Key management system provides cardinal security services such as freshness, secrecy, key synchronization, authentication, independence, ratification, and forward and backward secrecy [39][40]. Kerckhos‟ Principle stated that the attacker knows everything about the security solution [41] with the exception of the key.

6. *Trusted network*: Vulnerabilities may bring forth possibility for an attacker to penetrate a network, make headway admission to control software, hence make an alteration to load conditions to destabilize the grid in unpredictable ways. The lowest tone of a harmonic series of system-level components of a network security admitted trusted network [42]. Thus, approaches to secure networked technologies [43] and to protect privacy must be designed and implemented ahead in the transition to the smart grid.

7. *Interoperability and security*: The interoperability proffered by IP has enabled converged networks that provide both data and voice to become common in businesses [44], and a variety of triple play providers currently offer residential data, voice, and video on converged networks [45]. Interoperability is a primary or essential component of borderless smart grid networks, it is to be protected with solid and strong cyber security.

8. *Gap analysis*: Undiscovered information security gaps are able to be distinguished and palliated by integrating both electronic and physical information components [46][47].

9. *Reliable data storage system*: In accompaniment with the establishment of grid environments [48], various assurances were established virtually. These depict where the end-user is concerned over grid services that are hosted on high-end resources including high-priced instrumentation and data storage systems, connected via reliable and high-speed networks.

10. *Cybersecurity guidelines*: Some federal government agencies have developed, and are currently developing more security guidelines and best practices for smart grid [49]. Testimony argues that technology is growing progressively as potential instrument for terrorist organizations. This extend to the outgrowth of a new threat in the form of cyber terrorists. Cyber terrorists attack technological features such as the Internet in order to help foster their cause [50]. Hence, the nature of the responses in term of necessary to preserve the future security of our society become prior to action.

11. *Law enforcement*: When the public sharing of information about a specific location‟s energy used [51][52] is also a distinct possibility, law enforcement plays a mighty significant role, particularly when the concern exists that the prevalence of granular energy data could lead to actions on the part of law enforcement.

12. *Improved wireless technology*: Community benefits ease of cost savings [53], flexibility and deployment along with the establishment of wireless communications. However, with the existing of wire-line or wireless networks [54], immediate concern revolves around the security protections for easily accessible devices such as the meters and the related communications hardware. A man-in-the-middle attack exploites a wireless communication network armless with proper security protocols.

13. *Controlled power consumption*: Power consumption is affected by the program [55] running on a microprocessor [56]. Low and high-level language optimization techniques appear as an alternative in low power consumption analysis [57][58].

14. *Protect secret*: Protecting secret [59] has been identifies as one of the eight objectives of Commercial Information Systems Security. Some data and information are categorized secret and required unique or specific password login access to be obtained, read or copied where [60] access to these areas of the network resources needs authentication of the user nodes and the associated controls.

15. *Cryptographic protocols*: Cryptographic algorithms are necessitated to convert plaintext into ciphertext and vice versa [61]. The convertion of plaintext into ciphertext makes it impossible for an attacker to possess plaintext from a ciphertext without cognizing a key. It is a sequence of bits and serves as a parameter for transformation. In the area aiming to provide secure authentication and communication service negotiation in online presence notification systems [62], Cryptography practice is [63][64] beneficial to be employed.

16. *Encryption policies*: It is relatively well known that the encryption policies such as encrypting sensitive data that is either at rest in databases or in motion such as portable devices, emails and instant messages [65] can help to minimize the probability of insider"s misuse. Technical control against [66] insider attacks ought to be made up out of encryption.

Smart grid functional requirements, interoperability and cybersecurity standards are still evolving, besides aiming to prevent it from cyber-security threats, this set of information security functional requirements also play a significant role specifically to address security concerns that would arise from the transformation to a smart grid system. We should understand not only the technical but also the legal, policy and social requirements of cyber security.

# 7. FUTURE RESEARCH DIRECTION

The matching of consumer requirement and functional requirement of information security of the smart grid prevail an interesting sight for future research development.

When comes to discover the relationship between information security with customer trust and satisfaction, it is interesting to prove that there is a relationship between consumer requirement and the functional requirement of the information security criteria in a smart grid that would impact consumer"s trust and satisfaction towards the entire system, as in Table III below.

The future work intends to prove that there is a significant positive relationship between the said elements with the help of fuzzy logic value which would serve to eliminate ambiguity during data insertion and validation. Cx represents customer trust; Sx represents customer satisfaction; Iy represents information security; when the value of Sx and Cx on the Xs,c increase, the Iy also will increase at a different pace formula on Yi.

Table 3. Relationship Graph

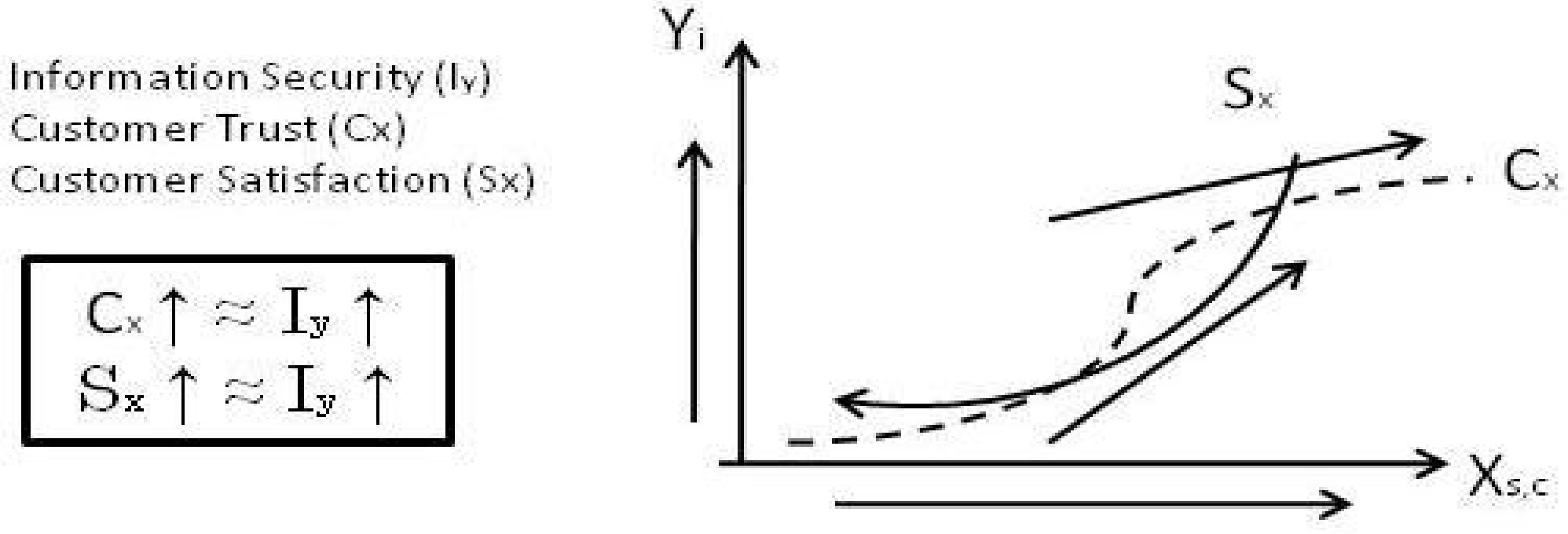

It takes aim to reveal the significant information security criteria in a smart grid that would impact consumer trust and satisfaction towards the entire system.

# 8. CONCLUSION

This paper review previous literature on smart grid information technology and the needs for smart grid information security. Then, the concept hermeneutic circle of the paper methodology is presented. The Hermeneutic Circle is indeed a good methodology applied in this paper as it gives the entire mapping of this study where the functional requirement could be identified one by one. Networking trend successfully elucidate the significant of information security as in Table 1 that leads to the generation idea of working on the collection of data regarding information security functional requirement identification showed detailed in Table 2.

Sixteen identified information security functional requirement were developed with the support of facts and references. This set of grid information security functional requirements defined the approach and serve as a base for the information security policy developers and utilities to refer as a basic functional requirement to satisfy consumer trust and satisfaction. Because a smart grid utilizes digital technology to provide two-way communication between suppliers and consumers" home electronics through the use of smart meters, information security protection measures need to be consumer friendly and practicable to be implemented on all level within a community or organization, making it more reliable, energy-efficient, and better able to serve all needs.

This paper has tutorial contents where some related backgrounds were provided, especially for networking community, covering the cyber security requirement of smart grid information infrastructure. It provides a methodology and some information security functional requirements conceptually as original contributions. This paper aims to contribute a sight for the readers to have a functional knowledge of the electric power grid and a better understanding of cyber security.

## ACKNOWLEDGEMENTS

The authors wish to thank Program GCOE, Graduate School of Science and Technology Meiji University and the Japanese Government Scholarship (MONBUKAGAKUSHO: MEXT) for the sponsor, Professor Mimura Masayasu and Professor Sugihara Kokichi from GCOE Program of Meiji University for the mentor. The authors also express gratitude to Dr. Zainol Mustafa and Dr. Rika Fatimah from School of Mathematical Sciences National University of Malaysia for discussion and ideas.

**Author 1**

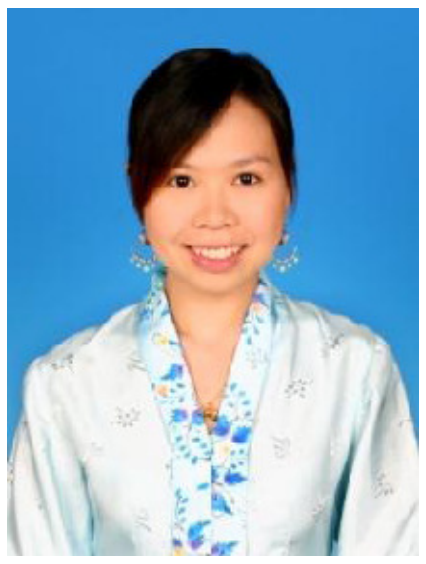

Amy Poh Ai Ling was a member of The National University of Malaysia (UKM) Fellowship stationed in Ibrahim Yaakub Residential College. Upon completing her Master in Sciences (MSc) in Quality and Productivity Improvement under Mathematical Sciences Department of UKM, she became a Parts Quality Assurance (Electronic) Engineer in EMCS Sony. Currently she is attaching with MIMS Ph.D. Program in Meiji University, Japan with MEXT scholarship awarded. The author always has an enthusiasm for mathematical quality and safety studies.

**Author 2**

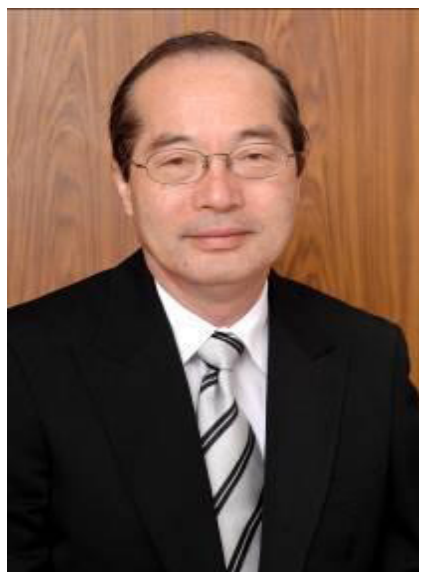

Mukaidono Masao served as a full-time lecturer at Faculty of Engineering, Department of Electrical Engineering in Meiji University from 1970. Even since then, he was promoted to Assistant Professor on 1973 and as a Professor on 1978. He contributed as a researcher in an Electronic Technical Laboratory of the Ministry of International Trade and Industry (1974), Institute of Mathematical Analysis of Kyoto University (1975) and as a visiting researcher at University of California in Berkeley (1979). He then became the Director of Computer Center (1986) and Director of Information Center (1988) in Meiji University. At present, he is a Professor and Dean of the School of Science & Technology, Meiji University. He is also the honourable Councillor of Meiji University